# The largest elephant in the room: aerosol masking

Guy R. McPherson[1], Beril Kallfelz-Sirmacek[2], William M. Kallfelz[3]
[1]School of Natural Resources and the Environment, University of Arizona, Tucson, AZ, USA
[2]Independent researcher, Overijssel, The Netherlands
[3]Department of Philosophy and Religious Studies, Mississippi State University, Mississippi, USA

**Abstract**
The aerosol masking effect, or "global dimming," is a well-documented instance of our ongoing climate predicament: Efforts to mitigate greenhouse gas production by curbing industrial activity or transitioning to "clean" sources such as solar or nuclear inevitably and inadvertently accelerate planetary warming. This undesired outcome results from aerosols that result from industrial activity. This article describes some climate mitigation strategies destined to fail because they do not account for the aerosol masking effect. Moreover, many political and ethical initiatives emphasize the "business as usual" framework, thus worsening the conundrum. We suggest instead a set of ethical and policy initiatives to address the challenge of aerosol masking, and we describe how our contemporary predicament can be ameliorated in the face of these significant and often-overlooked challenges.

## 1. Introduction

Aerosol masking effect is one of the radiative forces that cools Earth (Ångström, 1929). Earth is experiencing rapid warming, and a reduction of aerosol masking will very quickly contribute to additional warming. Aerosol masking is routinely ignored or underestimated by climate-mitigation strategies. We provide a brief review of relevant literature on the importance of the aerosol masking effect for slowing additional warming of the planet. We address the effectiveness of commonly touted mitigation strategies in light of the aerosol masking effect. In addition, we discuss the cost of ignoring the aerosol masking effect for industrial civilization and for life on Earth. We conclude by suggesting actions focused on environmental ethics and planetary hospice.

## 2. Background

Aerosols act as cloud condensation nuclei (CCN) to alter cloud properties and precipitation (Ångström, 1929), (Twomey, 1974), (Albrecht, 1989), thereby influencing the Earth's radiation budget and hence climate change. An increase in CCN creates a cloud with more droplets. Consequently, more solar radiation is reflected back to space, thus exerting a negative climate forcing. This is known as the cloud albedo effect or the Twomey effect (Twomey, 1974), and it has been reported since 1929 (Ångström, 1929). Despite this long history of description, aerosol masking remains rarely reported by media personalities and government officials, thereby obscuring from the public an important component of climate change. Masking from aerosols is commonly underestimated (Jia et al., 2021), which hampers the utility of climate models and therefore societal application of mitigation strategies.

Recent research reiterates the importance of accounting for aerosol masking in our attempts to understand the full impacts of anthropogenic forcing on climate change. COVID-19 "lockdown" periods led to loss of aerosol masking and a subsequent rise global temperature of up to 0.3 C, indicating that aerosol radiative forcing reductions are the largest contributor to surface temperature changes (Gettelman et al., 2021). The degree to which aerosols cool the earth has been grossly underestimated, necessitating a recalculation of climate-change models to more accurately predict the pace of global warming (ScienceDaily, 2019): "The fact that our planet is getting warmer even though aerosols are cooling it down at higher rates than previously thought brings us to a Catch-22 situation: Global efforts to improve air quality by developing cleaner fuels and burning less coal could end up harming our planet by reducing the number of aerosols in the atmosphere, and by doing so, diminishing aerosols' cooling ability to offset global warming." Loss of aerosol masking and the related temperature increase have been reported by many scientists at local and regional levels (Levy, 2013).

## 3. Climate mitigation strategies and the largest elephant in the room

In this section, we address the most commonly referenced climate mitigation strategies. We conclude that they will be ineffective or counterproductive as strategies to address ongoing, abrupt planetary warming.

### 3.1. Solar panels, windmills and other "renewable" energy resources

As fossil-fuel-powered industrial activity increased within the last two decades, greenhouse gases and aerosols also increased, as reported by the IPCC in AR6 (IPCC, 2021). The construction and implementation of "renewable energy" strategies such as solar panels and windmills cause reduction of the cooling aerosol gases, and thereby lead to an increased global temperature. In addition, industrial civilization is a heat engine, regardless how it is powered (Garrett et al., 2020), (Garrett, 2015), (Garrett, 2014), (Garrett, 2012(a)), (Garrett, 2012(b)), (Garrett, 2011). We do not support the continued use of fossil fuels. However, we do support full understanding of societal actions. Whereas mitigation might have limited the use of fossil fuels before the great acceleration began in the 1950s (Steffen et al., 2015), implementation of the same strategies today will cause rapid planetary heating due to a reduction in aerosol masking.

### 3.2. Nuclear power plants

Nuclear power plants accelerate heating associated with the loss of aerosol masking. In addition, the possible "meltdown" of nuclear facilities creates tremendous danger for all living species (Ehrlich, 1984). The release of ionizing radiation as nuclear facilities implode can destroy stratospheric ozone, therefore causing very rapid overheating (Tran, 2017), (Garcia-Sage, K., et. al., 2017).

### 3.3. Population control

Reduction of the human population will lead to the reduction of industrial activity. As with the previous example, population control could have been an important technique for mitigating climate change, had it been applied before the great acceleration. However, contemporary implementation of this approach will rapidly destroy human habitat, as described above.

### 3.4. Rewilding and Deforestation
Rewilding refers to the restoration of human-impacted ecosystems. This approach can be accompanied by the diversification of crops so that they are better able to survive changing climates. Although there have been a few local initiatives, it seems unlikely these actions will be implemented globally due to a profound societal desire to maintain industrial civilization as a means of pursuing economic growth. Even an effort to plant millions of trees will fail due to lack of space on Earth to plant the number of trees it would take to prevent the climate from additional, extreme warming (Velasco et al., 2016).

### 3.5. Migrating species
Moving animal and plant species to higher altitudes or higher latitudes will not guarantee retention of their habitat. In addition, a long history of introducing new species to any area with adverse consequences suggests that this approach is ill-advised (Garrett Hardin, 2009), (Kuhlenkamp et al., 2017).

### 3.6. Carbon capture storage and utilization
Carbon capture storage (CCS) facilities emit more carbon than they capture (Global witness, 2022). Since 2015, Shell's CCS has prevented the release of 5 million tonnes of carbon dioxide into the atmosphere, but it has also released a further 7.5 million tonnes. In other words, Shell's plant appears to have the same carbon footprint as 1.2 million gasoline-powered cars each year. Figures 1 and 2 demonstrate the greenhouse gas (GHG) input and output of Shell's CCS facility, respectively. The environmental footprint of CCS facilities is not included in these figures. The implementation of CCS facilities encourages investors and businesses to remain confident as they create an adverse environmental footprint, trusting that the GHG of their activities eventually will be compensated.

**Greenhouse gas emissions from Shell's Albertan fossil hydrogen plant, 2015-19**

|  | Source of GHGs | | | | | |
|---|---|---|---|---|---|---|
|  | Supply chain for fossil gas | On-site CO2 produced at the hydrogen plant | Electricity to power the CCS system | Total GHGs produced | Total GHGs captured | Total GHGs released |
| Tonnes of GHGs | 1,580,000 | 10,023,000 | 868,000 | 12,469,000 | 4,813,000 | 7,656,000 |
| % of overall GHGs | 12.7% | 80.3% | 7% | 100% | 38.6% | 61.4% |

Global Witness estimates, see Annex for the methodology and sources.

Figure 1 Detailed view of the GHG emissions from Shell's CCS facility between 2015 and 2019 (Global witness, 2022)

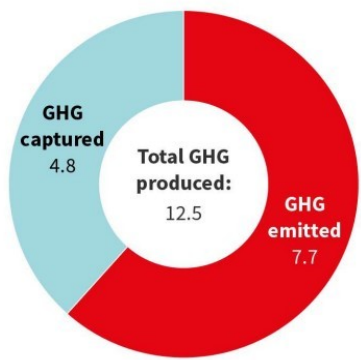

Figure 2  Percentage of the captured GHG and the created GHG emissions from Shell's CCS facility between 2015 and 2019 (Global witness, 2022).

### 3.7. Sustainable Development Goals (SDGs)
According to Cambridge Dictionary, the meaning of 'sustain' is to cause or allow something to continue for a 'period' of time. Therefore, the statement self-clarifies its short-term solution to allow the existing state. Needless to say, SDGs are designed for an unlimited growth scenario (also known as the "business-as-usual" scenario), which does not consider the limited resources of the planet. Thus, SDGs will not help overcome challenges of higher planetary temperatures that cause loss of basic materials such as food, drinking water, topsoil, and so on.

### 3.8. Trusting human or artificial intelligence
Artificial Intelligence (AI) is a mathematical tool used to create complex functions that can create maps, make future projections, find specific patterns, and explain correlations between different features. The contribution of advanced AI methods to automate such processes is unquestionable in terms of saving time for creating results and extracting knowledge from data that is difficult if not impossible to observe otherwise. However, the power of AI in representing highly complex functions must not lead to false expectations such as finding a solution to reverse climate change. After all, even the conservative Intergovernmental Panel on Climate Change (IPCC) concluded that climate change is abrupt (IPCC, 2018: Annex I) and irreversible (IPCC Special Report, 2019). In the former case, the IPCC indicated that, "even abrupt geophysical events do not approach current rates of human-driven change." In the latter case, the IPCC found that, "Ocean acidification and deoxygenation, ice sheet and glacier mass loss, and permafrost degradation are expected to be irreversible on time scales relevant to human societies and ecosystems."

### 3.9. Other complex mitigation strategies
Many scientists with good intentions have been working to design complex mitigation strategies that rely on a combination technologies that either do not exist or have already been demonstrated to be ineffective (Steffen et al., 2015), (Ludacer et al. 2018). Such mitigation strategies focus on creating nuclear, solar, and wind-based energy platforms, creating CCS facilities to store more carbon than is released, increasing ocean alkalinity to decrease acidification, and so on. As previously explained, solving global warming by replacing the existing energy resources with nuclear and "renewables" ignores the aerosol masking effect. Methods such as adding alkaline substances to the ocean remain untested. Side effects of such activities to marine life, water quality, and the biosphere have not been explored.

**4. Discussion**

**4.1. Conscious ignorance and covering of the aerosol masking effect**

We must assume that the aerosol masking effect is consciously ignored by many scientists. As explained by (Thunström, 2014), this behavior can be labeled 'strategic ignorance' for the purpose of avoiding feelings of guilt about the situation. (Schwartz, 2012) explained this behavior as 'willful ignorance' to avoid facing the hard truth that might cause strong, negative emotions.

The tendency of humans to continue living in a so-called 'business as usual' manner is often referred to as 'normalcy bias.' By definition, normalcy bias is a psychological state of denial people enter, often in the event of a disaster, as a result of which they underestimate the possibility of the disaster actually happening, as well as its effects on their life and property. This denial is based on the assumption that if the disaster has not occurred until now, it will never occur (Schwartz, 2012). These logical fallacies are described by (Bacon, 1620). Normalcy bias is a part of human nature, but it can pose a serious threat to life and property. If people are not prepared for an impending disaster, it is likely to result in this psychological phenomenon.

**4.2. Cost of ignoring the aerosol masking effect**

The costs of ignoring the aerosol masking effect go beyond avoidance of guilt, self-denial, and self- delusion. As we rapidly lose species from Earth in the midst of a Mass Extinction Event (Cowie et al., 2022) persistently spending time and resources on useless mitigation plans can only lead to loss of aerosols and increasingly rapid loss of habitat.

In addition to the loss of aerosol masking that will undoubtedly lead to loss of habitat for humans, there is a moral issue associated with ignorance. Just as the medical profession deemed it reasonable to withhold information from patients with a terminal diagnosis through the 1960s, many scientists, media personalities, and politicians have incorrectly determined that withholding information about abrupt, irreversible climate change is a moral approach (McPherson, 2019(a), McPherson, 2019(b)). Obviously, we disagree with this approach.

**4.3. Accumulated heat in the ocean**

The Earth Energy Imbalance (EEI) is the difference between the amount of energy from the sun arriving at the surface of Earth and the amount of energy returning to space. The energy accumulated in the ocean increases the large disparity and makes it impossible to equalize the EEI. (Loeb et al., 2021) reported that the difference between the incoming and the returning energy has already been doubled. Therefore, Earth keeps collecting more energy, even in light of the aerosol masking effect. However, losing aerosol masking due to an ineffective mitigation strategy will only contribute to the increase of this imbalance and therefore lead to an accelerated increase in the global-average temperature.

**4.4. Overlooked tipping points**

The Intergovernmental Panel on Climate Change concluded in its *IPCC Special Report on the Ocean and Cryosphere in a Changing Climate (IPCC Special report, 2019)* that, "Ocean acidification and deoxygenation, ice sheet and glacier mass loss, and permafrost degradation are expected to be irreversible on time scales relevant to human societies and ecosystems." This IPCC report indicates an overheated ocean was responsible for the irreversibility of climate change. The irreversibility of climate change comes from the fact that environmental "tipping points" have been overshot. A single self-reinforcing feedback loop, or "tipping point," is sufficient to demonstrate the irreversibility of climate change.

**4.5. Rate of Environmental Change**

To date, the geological and climatological records indicates that the Earth has witnessed seven mass extinction events (MEEs) before the present ongoing one (Corso et al. 2020). The present ongoing MEE (often referred to as the Great Anthropocene Extinction) differs significantly from previous events. The earlier MEEs occurred because of rapid environmental changes, although the rapidity of past changes was much slower than the ongoing event (IPCC, 2022) (McPherson et al., 2022). A global-average temperature increase of 5-6 C above the 1750 baseline within a few centuries will doom to extinction all life on Earth (Strona, et al., 2018), yet we are proceeding as if such a global-average rise in temperature will never occur despite historical warnings to the contrary. Continuing to follow a 'business as usual' scenario ensures further exponential changes will occur. However, use of "renewable" or nuclear energy will greatly accelerate the global-average rise in temperature due to loss of atmospheric aerosols.

**4.6. General Methodological Comments**

Prior to expanding on a broader discussion involving political and ethical ramifications, we briefly summarize the information presented above by calling attention to two major logical and methodological shortcomings:
- **The Reductionist Bias (RB):** Stated briefly, it is a logical fallacy resulting in the illicit inference: *X* essentially involves *Y* entails that *X* is nothing but *Y*. (Examples are plain enough to spot: For example, in neuroscience, tendentious conclusions of the materialistic sort are often drawn as follows:

'Consciousness essentially involves brain activity, therefore consciousness is nothing but brain activity.)
- **The Exclusivist-Isolationist Bias (EIB):** Following the heels of the Reductionist Bias, the EIB assumes that a negligible role is played by interactive effects of the system of study of interest.

Together these biases represent two key failures in much standard scientific methodology: the failure of adequate integrative thinking (an example of **RB**) and the inability to consider the behavior of large, complex systems (an example of the **EIB**). For instance, even in papers that address aerosol masking, interactive effects are ignored and trivialized (e.g., rapid warming precipitating inevitable tipping points), and synergistic consequences are not subject to scientific scrutiny.

**4.7. Environmental ethics and policy**

Because it appears that climate mitigation goals will not address climate change in time to preserve habitat for our species–after all, climate change is already abrupt and irreversible–we suggest using available time and resources to increase activities that provide improved living conditions for extant species. Consistent with McPherson's previous work (McPherson, 2019(b)), we propose planetary hospice. Ineffective policy results from a mismatch between the contexts of values and goals and the circumstantial factors with normative implications that demand a much different response than suggested by such inherently misguided norms blind to the evidence suggested by such circumstantial constraints. In extreme cases, this mismatch can result in **moral hazard**, a common pathology endemic within diverse organizational contexts. Moral hazard is best understood as a systemic breakdown of accountability and enforcement. It arises from structural and causal injustices and systemic oversights borne of logical and systemically methodological errors, as well as implicit biases mentioned above.

More precisely stated, moral hazard is best understood as a kind of "feedback" between (at least) t**hree parties** (whether individual agents or organizations). It is a particular and peculiarly pointed combination of the problem of diffusion of responsibility alongside disabling of accountability: **If an agent or party is incentivized to act in a particularly risk-prone fashion, they are liable to do so regardless of the consequences.** There are at least three parties are involved, as the "spillover effects" or costs (usually externalized) are often felt by parties not directly affected by the transactions between the two principal parties.

Cases of moral hazard often emerge in market-based economic systems characteristic of natural ebbs and flows. Consider (Figure 3) the "boom-bust" cycles marked by overproduction and overconfidence followed by the inevitable "corrections" or "lows" in the form of "shocks" to the system. When unsustainable bubbles are artificially prolonged and organizations deemed "too big to fail" are subsidized by agencies with oversight, this can incentivize risk-prone behavior. This activity can descend into a moral hazard: The broader social and environmental costs are inevitably externalized, or rendered invisible from scrutiny and accountability. The following schematic illustrates the "positive feedback" effect of moral hazard and demonstrates its detrimental consequences over time:

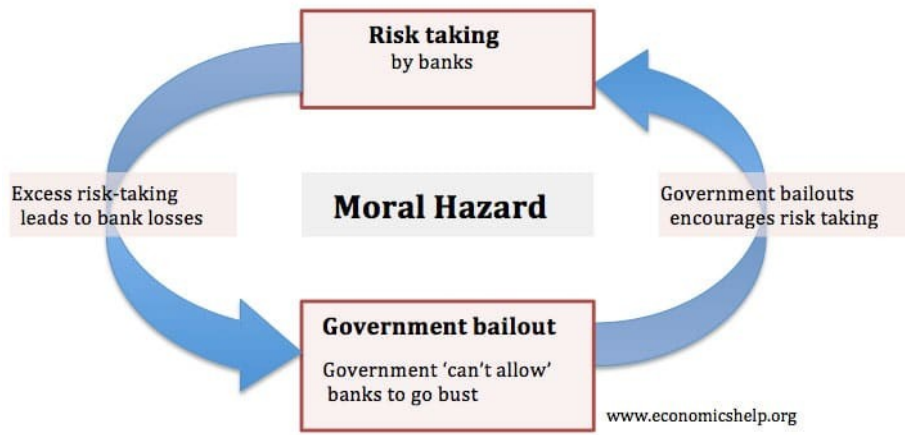

*Figure 3 The dynamics illustrating the incentives of bailouts serves as a generally instructive aid for understanding the mechanisms constituting moral hazard.*

The feedback of moral hazard endemic in contemporary climate policy (Figure 4) results from a complex psychological and social matrix of factors acting in concert, reinforcing the **Normalcy bias** as discussed in **Section 4.1.** These cognitive biases act against the framework required for responding ethically and effectively to the emergency of abrupt, irreversible climate change. They result from selective pressures favoring the local and the immediate over the global and the distal, in addition to assuming temporal continuity in the structure of anticipations and expectations (Gardiner, 2006). Powerful political, policy, and economic forces conspire to amplify such biases to the extent that the moral hazard is manifest in full form in a closed-loop causal feedback of pathology (Gardiner, 2006), further reinforcing this top-down application:

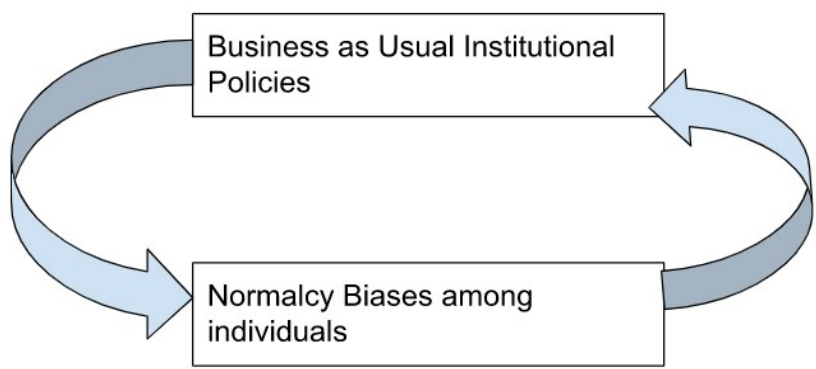

*Figure 4 Moral hazard in business as usual climate policies*

Planetary hospice, on the other hand, incentivizes individuals to take personal responsibility for their emotional, psychological, and moral well-being in ways that resonate with core humanitarian values (e.g., compassion, empathy, generosity, appreciative joy toward all sentient beings). These values encourage a generally Stoic disposition in the face of the collapse of industrial civilization and planetary ecosystems that are currently underway. This approach encourages attitudes of real courage and surrender of illusions. Moral hazard, on the other hand, inevitably results when the twin evils of denialism and utopianism take hold. We argue that much can still be done to foster stewardship and community support, but only in the face of an authentically courageous, reality-based frame of mind. Planetary hospice adopts and implements a set of values and norms to promote conscientious practices in pursuit of a holistic life.

## 5. Conclusions

Abundant peer-reviewed literature demonstrates that the most popularly recommended strategies to mitigate for ongoing, abrupt climate change cause Earth to heat up very abruptly due to loss of aerosol masking. We briefly suggest implementation of the idea of planetary hospice as a positive path forward at the levels of individuals, communities, and society.

## 6. Acknowledgements

Data sharing is not applicable to this article as no new data were created or analyzed in this study. The authors have no conflict of interest to declare.

## 7. Author statement

Authors G.R.M., B.K.S. and W.M.K. have participated equally and sufficiently in the conception and design of this work, as well as the writing of the manuscript, to take public responsibility for it.

We believe the manuscript represents valid work. We have reviewed the final version, and we approve it for publication. Neither this manuscript nor one with substantially similar content under my authorship has been published or is being considered for publication elsewhere.

## 8. References

Albrecht, B. A. (1989). Aerosols, cloud microphysics, and fractional cloudiness. Science 245, 1227–1230 (1989).

Ångström, A. (1929). On the Atmospheric Transmission of Sun Radiation and on Dust in the Air. Geografiska Annaler, 11:156–166.

Bacon, Sir (1620). Novum Organum. P. F. Collier.

Cowie, R.H., Bouchet, P. and Fontaine, B. (2022), The Sixth Mass Extinction: fact, fiction or speculation? Biol Rev, 97: 640-663. https://doi.org/10.1111/brv.12816

Dal Corso, J., Bernardi, M., Sun, Y., Song, H., Seyfullah, L. J., Preto, N., Gianolla, P., Ruffell, A., Kustatscher, E., Roghi, G., Merico, A., Hohn, S., Schmidt, A. R., Marzoli, A., Newton, R. J., Wignall, P. B., and Benton, M. J. (2020). Extinction and dawn of the modern world in the Carnian (Late Triassic). *Science advances*, *6*(38), eaba0099. https://doi.org/10.1126/sciadv.aba0099

Ehrlich, P., Sagan, C. (1984), The cold and the dark: the world after nuclear war. The conference on the long-term biological consequences of nuclear war. New York.

Fawzy, S., Osman, A.I., Doran, J. *et al.* (2020). Strategies for mitigation of climate change: a review. *Environ Chem Lett* 18, 2069–2094. https://doi.org/10.1007/s10311-020-01059-w
https://link.springer.com/article/10.1007/s10311-020-01059-w

Gettelman, A., Lamboll, R., Bardeen, C. G., Forster, P. M., & Watson-Parris, D. (2021). Climate impacts of COVID-19 induced emission changes. *Geophysical Research Letters*, 48, e2020GL091805. https://doi.org/10.1029/2020GL091805

Jia, H., Ma, X., Yu, F. *et al.* (2021). Significant underestimation of radiative forcing by aerosol–cloud interactions derived from satellite-based methods. *Nat Commun* 12, 3649 . https://doi.org/10.1038/s41467-021-23888-1

Levy, H., Horowitz, L. W., Schwarzkopf, M. D., Ming, Y., Golaz, J.-C., Naik, V., and Ramaswamy, V. (2013), The roles of aerosol direct and indirect effects in past and future climate change, *J. Geophys. Res. Atmos.*, 118, 4521–4532, doi:10.1002/jgrd.50192.

IPCC, 2018: Annex I: Glossary [Matthews, J.B.R. (ed.)]. In: Global Warming of 1.5°C. An IPCC Special Report on the impacts of global warming of 1.5°C above pre-industrial levels and related global greenhouse gas emission pathways, in the context of strengthening the global response to the threat of climate change, sustainable development, and efforts to eradicate poverty [Masson-Delmotte, V., P. Zhai, H.-O. Pörtner, D. Roberts, J. Skea, P.R. Shukla, A. Pirani, W. Moufouma-Okia, C. Péan, R. Pidcock, S. Connors, J.B.R. Matthews, Y. Chen, X. Zhou, M.I. Gomis, E. Lonnoy, T. Maycock, M. Tignor, and T. Waterfield (eds.)]. Cambridge University Press, Cambridge, UK and New York, NY, USA, pp. 541-562, doi:10.1017/9781009157940.008.

IPCC (2021). Climate Change 2021: The Physical Science Basis, Technical Summary
https://www.ipcc.ch/report/ar6/wg1/

Global witness, Report (2022). Hydrogen's Hidden Emissions.
https://www.globalwitness.org/en/campaigns/fossil-gas/shell-hydrogen-true-emissions/

ScienceDaily, 2019. The Hebrew University of Jerusalem. "We need to rethink everything we know about global warming: New calculations show scientists have grossly underestimated the effects of air pollution."

Steffen W, Broadgate W, Deutsch L, Gaffney O, Ludwig C. (2015). The trajectory of the Anthropocene: The Great Acceleration. *The Anthropocene Review*. 2(1):81-98. doi:10.1177/2053019614564785 https://journals.sagepub.com/doi/abs/10.1177/2053019614564785

Strona, G., Bradshaw, C.J.A. Co-extinctions annihilate planetary life during extreme environmental change. *Sci Rep* 8, 16724 (2018). https://doi.org/10.1038/s41598-018-35068-1 https://www.nature.com/articles/s41598-018-35068-1

Ludacer, R., Orwig, J. (2018) There's so much CO2 in the atmosphere that planting trees can no longer save us, Business Insider Netherlands. https://www.businessinsider.nl/so-much-co2-planting-trees-cant-save-us-2017-5/

Loeb, N. G., Johnson, G. C., Thorsen, T. J., Lyman, J. M., Rose, F. G., & Kato, S. (2021). Satellite and ocean data reveal marked increase in Earth's heating rate. *Geophysical Research Letters*, 48, e2021GL093047. https://doi.org/10.1029/2021GL093047

IPCC. 2022. AR6 Synthesis Report: Climate Change, 2022 https://www.ipcc.ch/report/sixth-assessment-report-cycle/

IPCC Special report (2019). Magnan, A.K., M. Garschagen, J.-P. Gattuso, J.E. Hay, N. Hilmi, E. Holland, F. Isla, G. Kofinas, I.J. Losada, J. Petzold, B. Ratter, T.Schuur, T. Tabe, and R. van de Wal, 2019: Cross-Chapter Box 9: Integrative Cross-Chapter Box on Low-Lying Islands and Coasts. In: IPCC Special Report on the Ocean and Cryosphere in a Changing Climate [H.-O. Pörtner, D.C. Roberts, V. Masson-Delmotte, P. Zhai, M. Tignor, E. Poloczanska, K. Mintenbeck, A. Alegría, M. Nicolai, A. Okem, J. Petzold, B. Rama, N.M. Weyer (eds.)]. Cambridge University Press, Cambridge, UK and New York, NY, USA, pp. 657-674. https://doi.org/10.1017/9781009157964.009.

McPherson, G., Sirmacek, B., Vinuesa, R. (2022). Environmental thresholds for mass-extinction events. *Results in Engineering.* Vol 13 March 2022. https://www.sciencedirect.com/science/article/pii/S2590123022000123

Gardiner, S. (2006). A Perfect Moral Storm: Climate Change, Intergenerational Ethics, and the Problem of Corruption. *Environmental Values* 15, n. 3: 397-413. Reprinted in Clowney D. & Mosto, P., eds. *Earthcare: An Anthology in Environmental Ethics.* Lanham, MD: Rowman & Littlefield, 2009: 347-360.

Garcia-Sage, K., et. al. (2017). On the Magnetic Protection of the Atmosphere of Proxima Centauri .*The Astrophysical Journal Letters.* Vole. 844 n. 1,Spring 2017. B https://iopscience.iop.org/article/10.3847/2041-8213/aa7eca

Garrett Hardin. 2009. The Tragedy of the Commons. *Science* 162 (1968): 1243-1248. Reprinted in Clowney, Reprinted in Clowney D. & Mosto, P., eds. *Earthcare: An Anthology in Environmental Ethics.* Lanham, MD: Rowman & Littlefield, 2009: 429-439.

Garrett, T. J., M. Grasselli, and S. Keen, 2020. Past world economic production constrains current energy demands: Persistent scaling with implications for economic growth and climate change mitigation. PLOS ONE, doi:10.1371/journal.pone.0237672.

Garrett, T. J., 2015. Long-run evolution of the global economy II: Hindcasts of innovation and growth. Earth Syst. Dynam. 6, 655–698.

Garrett, T. J., 2014. Long-run evolution of the global economy Part I: Physical basis. Earth's Future 2, 127–151, doi: 10.1002/2013EF000171.

Garrett, T. J., 2012 (a). Can we predict long-run economic growth? Retirement Management Journal 2, 53-61.

Garrett, T. J., 2012 (b). No way out? The double-bind in seeking global prosperity alongside mitigated climate change, Earth System Dynamics 3, 1-17, doi:10.5194/esd-3-1-2012.

Garrett, T. J., 2011. Are there basic physical constraints on future anthropogenic emissions of carbon dioxide? Climatic Change, 104, 437-455 doi:10.1007/s10584-009-9717-9.

Kuhlenkamp, R., King, B. (2017). Introduction of Non-indigenous Species Handbook on Marine Environment Protection pp 487–516C https://link.springer.com/chapter/10.1007/978-3-319-60156-4_25

McPherson, G. (2019(a)) Becoming hope-free: Parallels between death of individuals and extinction of homo sapiens. *Clinical Psychology Forum.* 317, May 2019. URL: https://guymcpherson.com/wp-content/uploads/2019/06/Clinical-Psychology-Forum-May-2019.pdf

McPherson, G. (2019(b)). Going Halfway: Climate Reports Ignore the Full Evidence, and Therapists Ignore Grief Recovery. *Clinical Psychology Forum.* 317, May 2019. https://guymcpherson.com/wp-content/uploads/2019/06/Clinical-Psychology-Forum-Sequel-DRAFT-1.pdf

Schwartz, S. (2012). Climate Change and Willful Ignorance. Explore (New York, N.Y.). 8. 268-70. 10.1016/j.explore.2012.06.008. https://www.sciencedirect.com/sdfe/pdf/download/eid/1-s2.0-S1550830712001334/first-page-pdf

Thunström, L., van 't Veld, K., Shogren, J. F., Nordström, J. (2014). On strategic ignorance of environmental harm and social norms, Dans Revue d'économie politique 2014/2 (Vol. 124), pages 195 à 214 https://www.cairn.info/revue-d-economie-politique-2014-2-page-195.html

Tran, L. (2017). Proxima Centauri's Radiation Would Wipe Out an Earth-Like Atmosphere on Proxima b. *NASA Goddard Space Flight Center.* August 1, 2017 https://scitechdaily.com/proxima-centauris-radiation-would-wipe-out-an-earth-like-atmosphere-on-proxima-b/

Twomey, S. (1974). Pollution and the planetary albedo. Atmos. Environ. 8, 1251–1256.

## 9. Figures

Figure 1 Detailed view of the GHG emissions from Shell's CCS facility between 2015 and 2019 (Global witness, 2022)

Figure 2 Percentage of the captured GHG and the created GHG emissions from Shell's CCS facility between 2015 and 2019 (Global witness, 2022)

Figure 3 The dynamics illustrating the incentives of bailouts serves as a generally instructive aid for understanding the mechanisms constituting moral hazard

Figure 4 Moral hazard in business as usual climate policies

## 10.     Research highlights

- The aerosol masking effect is one of the radiative forces that cools Earth. This paper reviews the literature on the importance of the aerosol masking effect for slowing additional warming of the planet.
- We address the effectiveness of commonly addressed climate mitigation strategies in light of the aerosol masking effect.
- We discuss the cost of ignoring the aerosol masking effect for industrial civilization and for life on Earth.

We suggest actions focused on environmental ethics and planetary hospice.